\documentclass[12pt]{article}
\usepackage{amsmath}

\newcommand{\bm}{\begin{multiline}}
\newcommand{\beq}{\begin{equation}}
\newcommand{\eeq}{\end{equation}}
\newcommand{\beqs}{\begin{eqnarray}}
\newcommand{\eeqs}{\end{eqnarray}}

\begin{document}

\thispagestyle{empty}


\begin{flushright}
hep-th/0312285\\
WATPHYS-TH03/09
\end{flushright}

\hfill{}

\hfill{}

\hfill{}

\vspace{32pt}

\begin{center}
\textbf{\Large Nuttier (A)dS Black Holes in Higher Dimensions} \\[0pt]

\vspace{48pt}

Robert Mann and Cristian Stelea

\vspace{12pt}

\textit{Department of Physics, Waterloo University, 200 University Avenue
West, Waterloo, Ontario, Canada, N2L 3G1}
\end{center}

\vspace{30pt}

\begin{abstract}

We construct new solutions of the vacuum Einstein field equations with
cosmological constant. These solutions describe spacetimes with non-trivial
topology that are asymptotically $dS$, $AdS$ or flat. For a negative
cosmological constant these solutions are $NUT$ charged generalizations of
the topological black hole solutions in higher dimensions. We also point out
the existence of such $NUT$ charged spacetimes in odd dimensions and we
explicitly construct such spaces in $5$ and $7$ dimensions. The existence of
such spacetimes with non-trivial topology is closely related to the
existence of the cosmological constant. Finally, we discuss the global
structure of such solutions and possible applications in string theory.
\end{abstract}

\setcounter{footnote}{0}

\newpage

\section{Introduction}

Ever since the seminal papers of Bekenstein and Hawking it has been known
that the entropy of a black-hole is proportional to the area of the horizon.
This relationship can be generalized to a wider class of spacetimes, namely
those whose Euclidean sections cannot be everywhere foliated by surfaces of
constant (Euclidean) time. \ These situations can occur if the Euclidean
spacetime has non-trivial topology: the inability to foliate the spacetime
leads to a breakdown of the concept of unitary Hamiltonian evolution, and
mixed states with entropy will arise \cite{Hawking,Hunter}. Spacetimes that
carry a NUT charge are in this broader class.

Intuitively the NUT charge corresponds to a magnetic type of mass. The first
solution in four dimensions describing such an object was presented in ref. %
\cite{Taub,NUT}. Although the Taub-NUT solution is not asymptotically flat ($%
AF$), it can be regarded as asymptotically locally flat ($ALF$). The
difference appears in the topology of the boundary at infinity. If we
consider as example of an $AF$ space the Euclidean version of the
Schwarzschild solution then the boundary at infinity is simply the product $%
S^{2}\times S^{1}$. By contrast, in the presence of a NUT charge, the
spacetime has as a boundary at infinity a twisted $S^{1}$ bundle over $S^{2}$%
. Only locally we can untwist the bundle structure to obtain the form of an $%
AF$ spacetime. The bundles at infinity are labelled by the first Chern
number, which is in fact proportional to the NUT charge \cite{Hawking}. The
presence of a NUT charge induces a so-called Misner singularity in the
metric, analogous to a `Dirac string' in electromagnetism \cite{Misner}.
This singularity is only a coordinate singularity and can be removed by
choosing appropriate coordinate patches. However, expunging this singularity
comes at a price: in general we must make coordinate identifications in the
spacetime that yield closed timelike curves in certain regions.

There are known extensions of the Taub-NUT solutions to the case when a
cosmological constant is present. In this case the asymptotic structure is
only locally de Sitter (for a positive cosmological constant) or anti-de
Sitter (for a negative cosmological constant) and we speak about
Taub-NUT-(a)dS solutions. In general, the Killing vector that corresponds to
the coordinate that parameterizes the fibre $S^{1}$ can have a
zero-dimensional fixed point set (we speak about a `NUT' solution in this
case) or a two-dimensional fixed point set (referred to as a `bolt'
solution).

Generalizations to higher dimensions follow closely the four-dimensional
case \cite{Bais,Page,Akbar,Robinson,Awad,Lorenzo}. In constructing these metrics the
idea is to regard the Taub-NUT space-time as a $U(1)$ fibration over a $2k$%
-dimensional base space endowed with an Einstein-K$\ddot{a}$hler metric $%
g_{B}$. Then the $(2k+2)$-dimensional Taub-NUT spacetime has the metric: 
\begin{equation}
F^{-1}(r)dr^{2}+(r^{2}+N^{2})g_{B}-F(r)(dt+A)^{2}  \label{I1}
\end{equation}%
where $t$ is the coordinate on the fibre $S^{1}$ and $A$ has a curvature $%
F=dA$, which is proportional to some covariantly constant 2-form. Here $N$
is the NUT charge and $F(r)$ is a function of $r$. The solution will
describe a `NUT' if the fixed point set of $e^{0}=dt+A$\ (i.e. the points
where $F(r)=0$) is less than $2k$-dimensional and a `bolt' if the fixed
point set is $2k$-dimensional. We can consider in the even-dimensional cases
circle fibrations over base spaces that can be factorized in the form $%
B=M_{1}\times \dots \times M_{k}$ where $M_{i}$ are two dimensional spaces
of constant curvature. In this case we can have a NUT charge $N_{i}$ for
every such two-dimensional factor and in the above ansatz we replace $%
(r^{2}+N^{2})g_{B}$ with the sum $\sum_{i}(r^{2}+N_{i}^{2})g_{M_{i}}$. In
particular, we can use the sphere $S^{2}$, the torus $T^{2}$ or the
hyperboloid $H^{2}$ as factor spaces.

Note that the factor in front $g_{B}$\ of is never zero unless we go to the
Euclidean section. Hence when we shall consider the possible singularities
of the above metrics we shall focus mainly on their Euclidean sections,
recognizing that the Lorentzian versions are singularity-free (apart from
quasi-regular singularities \cite{Konk}), i.e. scalar curvature
singularities have the possibility of manifesting themselves only in the
Euclidean sections.

In this paper we generalize the above construction to odd-dimensional
space-times. We use a base space that is odd-dimensional\thinspace\ and
construct an $S^{1}$ bundle over an even-dimensional K$\ddot{a}$hler space $%
M $ that is a factor of the odd dimensional base space. Specifically we
assume that the base space can be factorized in the form $B=M\times Y$ and
employ the following ansatz for the metric of the odd-dimensional Taub-NUT
spaces: 
\begin{equation}
F^{-1}(r)dr^{2}+(r^{2}+N^{2})g_{M}+r^{2}g_{Y}-F(r)(dt+A)^{2}  \label{I2}
\end{equation}%
Here $g_{M}$ is the metric on the even-dimensional space $M$ while $g_{Y}$
is the metric on the remaining factor space $Y$. We explicitly construct
NUT-charged spaces in $3$, $4$, $5$, $6$ and $7$ dimensions. These solutions
represent new generalizations of the spacetimes studied in refs. \cite%
{Bais,Robinson,Awad,Lorenzo}.

Our conventions are: $(-,+,...,+)$ for the (Lorentzian) signature of the
metric; in $d$ dimensions our metrics will be solutions of the vacuum
Einstein field equations with cosmological constant $\lambda =\pm \frac{%
(d-1)(d-2)}{2l^{2}}$, which can be expressed in the form $G_{ij}+\lambda
g_{ij}=0$ or in the equivalent form $R_{ij}\pm\frac{d-1}{l^{2}}g_{ij}=0$.

\section{Taub-NUT-dS metrics in $3$ dimensions}

The only consistent way to construct a three-dimensional spacetime as a $U(1)
$ fibration over a two-dimensional base space with constant curvature is to
use as the base space an hyperboloid $H^{2}$. In this case we obtain the
following NUT-charged metric which is a solution of the vacuum Einstein
field equations with negative cosmological constant $\lambda =-\frac{1}{l^{2}%
}$: 
\begin{equation}
ds^{2}=-\frac{l^{2}}{16n^{2}}(dt+2n\cosh rd\theta )^{2}+\frac{l^{2}}{4}%
(dr^{2}+\sinh ^{2}rd\theta ^{2})  \label{H2}
\end{equation}%
where $n$ is the NUT charge. The signature of this metric is $(-,+,-)$. \ If
we require a Lorentzian signature $(-,+,+)$ then we must restrict the values
of the cosmological constant to be positive ($\lambda =\frac{1}{l^{2}}$) and
so we must analytically continue $l\rightarrow il$ in (\ref{H2}). Two other
solutions are given by: 
\begin{equation}
ds^{2}=-\frac{l^{2}}{16n^{2}}(dt+2n\sinh rd\theta )^{2}+\frac{l^{2}}{4}%
(dr^{2}+\cosh ^{2}rd\theta ^{2})  \label{32}
\end{equation}%
\begin{equation}
ds^{2}=-\frac{l^{2}}{16n^{2}}(dt+2ne^{r}d\theta )^{2}+\frac{l^{2}}{4}%
(dr^{2}+e^{2r}d\theta ^{2})  \label{33}
\end{equation}

If we analytically continue the coordinates $t\rightarrow i\chi $, $\theta
\rightarrow i\theta $, $r\rightarrow it$ we obtain a time-dependent metric
of the form: 
\begin{equation}
ds^{2}=\frac{l^{2}}{16n^{2}}(d\chi +2n\cos td\theta )^{2}+\frac{l^{2}}{4}%
(-dt^{2}+\sin ^{2}td\theta ^{2})  \label{34}
\end{equation}%
Another possibility is to analytically continue $\theta \rightarrow it$ and $%
n\rightarrow -in$ in (\ref{H2}) to obtain (with $t\rightarrow \theta $ and $%
\lambda <0$): 
\begin{equation}
ds^{2}=\frac{l^{2}}{16n^{2}}(d\theta +2n\cosh rdt)^{2}+\frac{l^{2}}{4}%
(dr^{2}-\sinh ^{2}rdt^{2})  \label{35}
\end{equation}%
\begin{equation}
ds^{2}=\frac{l^{2}}{16n^{2}}(d\theta +2n\sinh rdt)^{2}+\frac{l^{2}}{4}%
(dr^{2}-\cosh ^{2}rdt^{2})  \label{36}
\end{equation}%
\begin{equation}
ds^{2}=\frac{l^{2}}{16n^{2}}(d\theta +2ne^{r}dt)^{2}+\frac{l^{2}}{4}%
(dr^{2}-e^{2r}dt^{2})  \label{37}
\end{equation}%
which are all solutions of the Einstein field equations with negative
cosmological constant.

The above metrics correspond to the Lorentzian versions of some the
so-called Thurston geometries \cite{thurston}. These geometries are so named
because of Thurston's conjecture that a 3-manifold with a given topology can
be decomposed into a connected sum of simple 3-manifolds, each of which
admits one of eight geometries: $H^{3}$, $S^{3}$, $E^{3}$, $S^{2}\times
S^{1} $, $H^{2}\times S^{1}$, $Sol$, $Nil$ and $SL(2,R)$. In our case we can
apply a T-duality along the Hopf $S^{1}$ direction and untwist the circle
fibration to the product space $H^{2}\times S^{1}$ %
\cite{ads3}. Hence the spacetimes that we have obtained are T-dual to some
of the eight Thurston geometries.

\section{Taub-NUT-AdS/dS-like metrics in $4$ dimensions}

In four dimensions we can use as base spaces any Einstein metric. For
simplicity we shall consider the following cases: the sphere $S^{2}$, the
torus $T^{2}$ and the hyperboloid $H^{2}$. These metrics are solutions of
the vacuum Einstein field equations with a cosmological constant $\lambda =-%
\frac{3}{l^{2}}$ in $4$ dimensions and their rotating versions have been
presented and discussed in \cite{ortin}. We shall refer them as topological
Taub-NUT-AdS/dS spacetimes since in general the base manifold will be a
compact space that is not simply connected.

\begin{itemize}
\item $U(1)$ fibration over $S^2$.

The metric is given by: 
\begin{equation}
ds^{2}=-F(r)(dt-2n\cos \theta d\phi
)^{2}+F^{-1}(r)dr^{2}+(r^{2}+n^{2})d\Omega ^{2}  \label{41}
\end{equation}%
where $d\Omega ^{2}=d\theta ^{2}+\sin ^{2}\theta d\phi ^{2}$ is the metric
on the sphere $S^{2}$ and 
\begin{equation}
F(r)=\frac{r^{4}+(l^{2}+6n^{2})r^{2}-2mrl^{2}-n^{2}(l^{2}-3n^{2})}{%
l^{2}(n^{2}+r^{2})}  \label{42}
\end{equation}%
Notice that when the NUT charge $n=0$ we recover the Schwarzschild-AdS/dS
solution with 
\begin{equation}
F(r,n=0)=1-\frac{2m}{r}-\frac{r^{2}}{l^2}  \label{43}
\end{equation}

\item $U(1)$ fibration over $T^2$.

In this case we use as base space the torus $T^{2}$. The toroidal
Taub-NUT-AdS/dS solution is given by: 
\begin{equation}
ds^{2}=-F(r)(dt-2n\theta d\phi )^{2}+F^{-1}(r)dr^{2}+(r^{2}+n^{2})(d\theta
^{2}+d\phi ^{2})  \label{44}
\end{equation}%
Notice that in this case $A=2n\theta d\phi $ for which the curvature 2-form
is proportional with the volume element of the torus. The function $F(r)$ is
given by: 
\begin{equation}
F(r)=\frac{r^{4}+6n^{2}r^{2}-2ml^{2}r-3n^{4}}{l^{2}(n^{2}+r^{2})}  \label{45}
\end{equation}%
When $n=0$ we recover the toroidal Schwarzschild-AdS/dS solution, with: 
\begin{equation}
F(r,n=0)=\frac{r^{2}}{l^{2}}-\frac{2m}{r}  \label{46}
\end{equation}

\item $U(1)$ fibration over $H^{2}$.

The solution is given by: 
\begin{equation}
ds^{2}=-F(r)(dt-2n\cosh \theta d\phi
)^{2}+F^{-1}(r)dr^{2}+(r^{2}+n^{2})(d\theta ^{2}+\sinh ^{2}\theta d\phi ^{2})
\label{47}
\end{equation}%
where: 
\begin{equation}
F(r)=\frac{r^{4}+(6n^{2}-l^{2})r^{2}-2ml^{2}r-n^{2}(3n^{2}-l^{2})}{%
l^{2}(n^{2}+r^{2})}  \label{48}
\end{equation}%
Notice that for $n=0$ we recover the hyperbolic Schwarzschild-AdS/dS
solution, for which: 
\begin{equation}
F(r,n=0)=-1-\frac{2m}{r}-\frac{r^{2}}{l^2}  \label{49}
\end{equation}
\end{itemize}

Here $m$ is a mass parameter, and the spaces are realized as non-trivial
fibrations over a compact two-dimensional space. As shown in \cite{Chamblin}
one can have NUT and bolt solutions, with the exception of the one that
corresponds to a circle fibration over $H^{2}$, in which case there are no
NUT solutions.

\section{Taub-Nut-dS/AdS spacetimes in $5$ dimensions}

In even-dimensions the usual Taub-NUT construction corresponds to a $U(1)$%
-fibration over an even-dimensional Einstein space used as the base space.
Since obviously this cannot be done in odd-dimensions, we must modify our
metric ansatz in such a way that we can realize the $U(1)$-fibration as a
fibration over an even dimensional subspace of the odd dimensional base
space. In five dimensions our base space is three dimensional and we shall
construct the NUT space as a partial fibration over a two-dimensional space
of constant curvature. The spacetimes that we obtain are not trivial in the
sense that we cannot set the NUT charge and/or the cosmological constant
(now $\lambda =\frac{6}{l^{2}}$) to vanish.

Consider first a fibration over $S^{2}$. The ansatz that we shall use in the
construction of these spaces is the following: 
\begin{equation}
ds^{2}=-F(r)(dt-2n\cos \theta d\phi
)^{2}+F^{-1}(r)dr^{2}+(r^{2}+n^{2})(d\theta ^{2}+\sin ^{2}\theta d\phi
^{2})+r^{2}dy^{2}  \label{51}
\end{equation}%
The above metric is a solution of the Einstein field equations with
cosmological constant $\lambda $ provided%
\begin{equation}
F(r)=\frac{4ml^{2}-r^{4}-2n^{2}r^{2}}{l^{2}(r^{2}+n^{2})}  \label{52}
\end{equation}%
and%
\begin{equation}
n^{2}=\frac{l^{2}}{4}  \label{53}
\end{equation}

Let us consider next the Euclidean section of the above solution (obtained
by making the analytical continuations $t\rightarrow i\chi $ and $%
n\rightarrow in$): 
\begin{equation}
ds^{2}=F_{E}(r)(d\chi -2n\cos \theta d\phi
)^{2}+F_{E}^{-1}(r)dr^{2}+(r^{2}-n^{2})(d\theta ^{2}+\sin ^{2}\theta d\phi
^{2})+r^{2}dz^{2}  \label{53a}
\end{equation}%
where 
\begin{equation}
F_{E}(r)=\frac{r^{4}-2n^{2}r^{2}+4ml^{2}}{l^{2}(r^{2}-n^{2})}  \label{53b}
\end{equation}%
and the constraint $\lambda n^{2}=-\frac{3}{2}$ holds. Since we analytically
continue $n$ we must also analytically continue $l\rightarrow il$ for
consistency with the initial constraint on $\lambda $ and $n^{2}$.

In order to get rid of the usual Misner type singularity in the metric we
have to assume that the coordinate $\chi $ is periodic with period $\beta $.
Notice that for $r=n$ the fixed point of the Killing vector $\frac{\partial 
}{\partial \chi }$ is one dimensional and we shall refer to such a solution
as being a NUT solution. However, for $r=r_{b}$, where $r_{b}>n$ is the
largest root of $F_{E}(r)$, the fixed point set is three-dimensional and we
shall refer to such solutions as bolt solutions. Note that for either
situation the the period of $\chi $ must be $\beta =8\pi n$ to ensure the
absence of the Dirac-Misner string singularity.

In order to have a regular NUT solution we have to ensure the following
additional conditions:

\begin{itemize}
\item $F_{E}(r=n)=0$ in order to ensure that the fixed point of the Killing
vector $\frac{\partial }{\partial \chi }$ is one-dimensional.

\item $\beta F_{E}^{\prime }(r=n)=4\pi k$ (where $k$ is an integer) in order
to avoid the presence of conical singularities at $r=n$ (in other words, the
periodicity of $\chi $ must be an integer multiple of the periodicity
required for regularity in the $\left( \chi ,r\right) $ section; we identify 
$k$ points on the circle described by $\chi $).
\end{itemize}

It is easy to see that the above conditions lead to $k=1$ and $m_n=\frac{%
n^{4}}{4l^{2}}=\frac{l^{2}}{64}$. It is precisely for this value of the
parameter $m$ that the above solution becomes the Euclidean $AdS$ spacetime
in five-dimensions.

Let us now turn to the regularity conditions that we have to impose in order
to obtain the bolt solutions. In order to have a regular bolt at $r=r_{b}$
we have to satisfy similar conditions as before, with $r_{b}>n$:

\begin{itemize}
\item $F_E(r=r_b)=0$

\item $\beta F_{E}^{\prime }(r=r_{b})=4\pi k$ where $k$ is an integer.
\end{itemize}

The above conditions lead to $r_{b}=\frac{kn}{2}$ and 
\begin{equation}
m=m_{b}=-\frac{k^{2}l^{2}(k^{2}-8)}{1024}  \label{53c}
\end{equation}%
To ensure that $r_{b}>n$ we have to take $k\geq 3$; as a consequence the
curvature singularity at $r=n$ is avoided. We obtain the following family of
bolt solutions, indexed by the integer $k$: 
\begin{equation}
ds^{2}=F_{E}(r,k)(d\chi -2n\cos \theta d\phi
)^{2}+F_{E}^{-1}(r,k)dr^{2}+(r^{2}-n^{2})(d\theta ^{2}+\sin ^{2}\theta d\phi
^{2})+r^{2}dz^{2}  \label{53d}
\end{equation}%
where 
\begin{equation}
F_{E}(r,k)=\frac{256r^{4}-128l^{2}r^{2}-k^{2}l^{4}(k^{2}-8)}{%
256l^{2}(r^{2}-n^{2})}  \label{53f}
\end{equation}%
and $n=\frac{l}{2}$. One can check directly that the bolt solution is not
simply the AdS space in disguise by computing the curvature tensor of the
bolt metric and comparing it with that of the Euclidean AdS space.

We can obtain NUT spaces with non-trivial topology if we make partial base
fibrations over a two-dimensional torus $T^{2}$ or over the hyperboloid $%
H^{2}$. We obtain 
\begin{equation}
ds^{2}=-F(r)(dt-2n\theta d\phi )^{2}+F^{-1}(r)dr^{2}+(r^{2}+n^{2})(d\theta
^{2}+d\phi ^{2})+r^{2}dy^{2}  \label{54}
\end{equation}%
for the torus, where 
\begin{equation}
F(r)=\frac{4ml^{2}+r^{4}+2n^{2}r^{2}}{l^{2}(r^{2}+n^{2})}  \label{55}
\end{equation}%
where now the constraint equation takes the form $\lambda n=0$ where $%
\lambda =-\frac{6}{l^{2}}$; we can have consistent Taub-NUT spaces with
toroidal topology if and only if the cosmological constant vanishes. The
Euclidean version of this solution, obtained by analytic continuation of the
coordinate $t\rightarrow it$ and of the parameter $n\rightarrow in$ has a
curvature singularity at $r=n$. Note that if we consider $n=0$ in the above
constraint we obtain the AdS/dS black hole solution in five dimensions with
toroidal topology.

If the cosmological constant vanishes then we can have $n\neq 0$ and we
obtain the following form of the metric 
\begin{equation}
ds^{2}=-F(r)(dt-2n\theta d\phi )^{2}+F^{-1}(r)dr^{2}+(r^{2}+n^{2})(d\theta
^{2}+d\phi ^{2})+r^{2}dy^{2}  \label{55a}
\end{equation}%
where 
\begin{equation}
F(r)=\frac{4m}{r^{2}+n^{2}}
\end{equation}%
The asymptotic structure of the above metric is given by 
\begin{equation}
ds^{2}=\frac{4m}{r^{2}}(dt-2n\theta d\phi )^{2}+\frac{r^{2}}{4m}%
dr^{2}+r^{2}(d\theta ^{2}+d\phi ^{2}+dy^{2})
\end{equation}%
If $y$ is an angular coordinate then the angular part of the metric
parameterizes a three torus. The Euclidian section of the solution described
by (\ref{55a}) is not asymptotically flat and has a curvature singularity
localized at $r=0$. However, let us notice that for $r\leq n$ the signature
of the space becomes completely unphysical. Hence, for the Euclidian
section, we should restrict the values of the radial coordinate such that $%
r\geq n$.

In the case of a fibration over the hyperboloid $H^{2}$ we obtain: 
\begin{equation}
ds^{2}=-F(r)(dt-2n\cosh \theta d\phi
)^{2}+F^{-1}(r)dr^{2}+(r^{2}+n^{2})(d\theta ^{2}+\sinh ^{2}\theta d\phi
^{2})+r^{2}dy^{2}  \label{56}
\end{equation}%
where now $\lambda =-\frac{6}{l^{2}}$, 
\begin{equation}
F(r)=\frac{r^{4}+2n^{2}r^{2}-4ml^{2}}{l^{2}(r^{2}+n^{2})}  \label{57}
\end{equation}%
and the constraint $n^{2}=\frac{l^{2}}{4}$ holds.

The Euclidean section of these spaces is described by the metric 
\begin{equation}
ds^{2}=F_{E}(r)(dt-2n\cosh \theta d\phi
)^{2}+F_{E}^{-1}(r)dr^{2}+(r^{2}-n^{2})(d\theta ^{2}+\sinh ^{2}\theta d\phi
^{2})+r^{2}dy^{2}  \label{58}
\end{equation}%
where $n^{2}=\frac{l^{2}}{4}$ and 
\begin{equation}
F_{E}(r)=-\frac{r^{4}-2n^{2}r^{2}+4ml^{2}}{l^{2}(r^{2}-n^{2})}  \label{59}
\end{equation}%
and it is a Euclidean solution of the vacuum Einstein field equations with
positive cosmological constant. The coordinates $\theta $ and $\phi $
parameterize a hyperboloid, which after performing appropriate
identifications becomes a surface of any genus higher than $1$. In general
the metric has a curvature singularity located at $r=n=\frac{l}{2}$ with the
exception of the case in which $m_{n}=-\frac{l^{2}}{64}$ when the space is
actually the five-dimensional Euclidean $dS$ space in disguise.

In order to discuss the possible singularities in the metric first let us
notice the absence of Misner strings, the fibration over the hyperbolic
space being trivial in this case. Moreover, if we impose the condition that
there are no conical singularities at $r=r_{p}$, where $r_{p}$ is the
biggest root of $F_{E}(r)$, then we must set the periodicity $\beta $ of the
coordinate $\chi $ to be $\frac{4\pi }{|F_{E}^{\prime }(r=r_{p})|}$ . If we
take $r_{p}=n$ we obtain $\beta =8\pi n$ and $m=-\frac{l^{2}}{64}$, which
means that the NUT solution is the $dS$ space in disguise.

In order to determine the bolt solution one has to satisfy the following
conditions:

\begin{itemize}
\item $F_E(r=r_b)=0$

\item $\frac{4\pi }{|F_{E}^{\prime }(r_{b})|}=\frac{8\pi n}{k}$ where $k$ is
an integer and the period of $\chi $ is now given by $\beta =\frac{8\pi n}{k}
$; again we identify $k$ points on the circle described by $\chi $.
\end{itemize}

The above conditions lead to $r_{b}=\frac{kn}{2}$ and 
\begin{equation}
m=m_{b}=\frac{k^{2}l^{2}(k^{2}-8)}{1024}  \label{60}
\end{equation}%
We must take $k\geq 3$ to ensure that $r_{b}>n$, which again avoids the
curvature singularity at $r=n$. We obtain the following family of bolt
solutions, indexed by the integer $k$: 
\begin{equation}
ds^{2}=F_{E}(r)(d\chi -2n\cosh \theta d\phi
)^{2}+F_{E}^{-1}(r)dr^{2}+(r^{2}-n^{2})(d\theta ^{2}+\sinh ^{2}\theta d\phi
^{2})+r^{2}dz^{2}  \label{60a}
\end{equation}%
where 
\begin{equation*}
F_{E}(r)=\frac{-256r^{4}+512n^{2}r^{2}+k^{2}l^{4}(k^{2}-8)}{%
256l^{2}(r^{2}-n^{2})}
\end{equation*}%
and $n=\frac{l}{2}$.

\section{Taub-NUT-AdS metrics in $6$ dimensions}

In this section we shall describe Taub-NUT-like solutions for the vacuum
Einstein field equations with cosmological constant. In $6$-dimensions the
base space that we can use is $4$-dimensional and we shall use all the
possible combinations of products of $S^{2}$, $T^{2}$ and $H^{2}$.

\subsection{`Full' fibrations over the product base space}

\begin{itemize}
\item $U(1)$ fibration over $S^2\times S^2$

The metric is given by: 
\begin{eqnarray}
ds^{2} &=&-F(r)(dt-2n_{1}\cos \theta _{1}d\phi _{1}-2n_{2}\cos \theta
_{2}d\phi _{2})^{2}+F^{-1}(r)dr^{2}  \notag \\
&&+(r^{2}+n_{1}^{2})(d\theta _{1}^{2}+\sin ^{2}\theta _{1}d\phi
_{1}^{2})+(r^{2}+n_{2}^{2})(d\theta _{2}^{2}+\sin ^{2}\theta _{2}d\phi
_{2}^{2})  \notag \\
&&  \label{61}
\end{eqnarray}%
where: 
\begin{eqnarray}
F(r) &=&\frac{%
3r^{6}+(l^{2}+5n_{2}^{2}+10n_{1}^{2})r^{4}+3(n_{2}^{2}l^{2}+10n_{1}^{2}n_{2}^{2}+n_{1}^{2}l^{2}+5n_{1}^{4})r^{2}%
}{3(r^{2}+n_{1}^{2})(r^{2}+n_{2}^{2})l^{2}}  \notag \\
&&+\frac{6ml^{2}r-3n_{1}^{2}n_{2}^{2}(l^{2}+5n_{1}^{2})}{%
3(r^{2}+n_{1}^{2})(r^{2}+n_{2}^{2})l^{2}}  \label{62}
\end{eqnarray}%
Here the above metric is a solution of vacuum Einstein field equations with
cosmological constant ($\lambda =-\frac{10}{l^{2}}$) if and only if: 
\begin{equation}
(n_{1}^{2}-n_{2}^{2})\lambda =0  \label{63}
\end{equation}%
Consequently we see that differing values for $n_{1}$ and $n_{2}$ are
possible only if the cosmological constant vanishes.
\end{itemize}

In order to analyze the possible singularities of these spacetimes we shall
consider the corresponding Euclidean sections, obtained by analytic
continuation of the coordinate $t\rightarrow i\chi $ and of the parameters $%
n_{j}\rightarrow in_{j}$ where $j=1,2$.

If the cosmological constant is zero then one can have two distinct values
of the parameters $n_{1}$ and $n_{2}$. Let us assume first that $%
n_{1}=n_{2}=n$. Then the NUT solutions correspond to the fixed-point set of $%
\frac{\partial }{\partial \chi }$ at $r=n$ and we obtain the value of the
mass parameter to be $m_n=\frac{4n^{3}}{3}$ while the periodicity of the
coordinate $\chi $ is $12\pi n$. However, even if we avoid the conical
singularity at $r=n$ we still have a curvature singularity located at $r=n$,
as one can check by computing some of the curvature invariants.

For the bolt solution we shall still impose the periodicity of the
coordinate $\chi $ to be $\frac{12\pi n}{k}$, while the fixed-point set is
four-dimensional and located at $r=r_{b}=\frac{3n}{k}$; since $r_{b}>n$,
this implies that only the values $k=1,2$ are relevant. The mass parameter
is given by $m=m_{b}=\frac{n^{3}(k^{4}+18k^{2}-27)}{6k^{3}}$.

In the case in which the values of the two NUT charges $n_{1}$ and $n_{2}$
are different, one can assume without loss of generality that $n_{1}>n_{2}$.
In this case in the Euclidean section the radius $r$ cannot be smaller than $%
n_{1}$ or the signature of the spacetime will change. Generically, there is
a curvature singularity located at $r=n_{1}$; however for a certain value of
the mass parameter $m$ this curvature singularity is removed. The NUT
solution in this case corresponds to a two-dimensional fixed-point set
located at $r=n_{1}$. Removal of the Dirac string singularity forces the
periodicity of the coordinate $\chi $ to be $8\pi n_{1}$. The value of the
mass parameter is $m=m_{b}=\frac{n_{1}^{3}+3n_{1}n_{2}^{2}}{3}$, and only
for this value of the parameter $m$ the metric is well-behaved at $r=n_{1}$.
This is a similar situation to that in the five-dimensional case; the
curvature singularity at $r=n$ disappears if and only if $m$ is a specific
function of the NUT charge. For any other values of $m$ (as in the bolt
solutions) the curvature singularity is still present.

The bolt solution corresponds to a four-dimensional fixed-point set located
at $r=r_{b}=\frac{2n_{1}}{k}$, for which the periodicity of the coordinate $%
\chi $ is given by $\frac{8\pi n_{1}}{k}$ and the value of the mass
parameter is $m=m_{b}=\frac{n_{1}(12n_{2}^{2}-4n_{1}^{2})}{12}$. Since $%
r_{b}>n_{1}$, we must choose $k=1$, thereby avoiding the curvature
singularity.

If the cosmological constant is non-zero then the above constraint equation
will impose the condition $n_{1}=n_{2}=n$. The Euclidean section of the
metric has a curvature singularity located at $r=n$. The NUT solution
corresponds to a zero-dimensional fixed-point set of the vector $\frac{%
\partial }{\partial \chi }$ located at $r=n$. The periodicity of the
coordinate $\chi $ is given by $12\pi n$ while the value of the mass
parameter is $m=m_{b}=\frac{4n^{3}(l^{2}-6n^{2})}{3l^{2}}$ \cite%
{Awad,Lorenzo}. The bolt solution corresponds to a four-dimensional
fixed-point set at 
\begin{equation*}
r=r_{b}=\frac{1}{30n}(l^{2}\pm \sqrt{l^{4}-180n^{2}l^{2}+900n^{4}})
\end{equation*}%
while the mass parameter is given by: 
\begin{equation*}
m_{b}=\frac{%
-3r_{b}^{6}+(15n^{2}-l^{2})r_{b}^{4}+3n^{2}(2l^{2}-15n^{2})r_{b}^{2}+3n^{4}(l^{2}-5n^{2})%
}{6l^{2}r_{b}}
\end{equation*}%
In order to avoid the singularity at $r=n$ we shall impose the condition $%
r_{b}>n$ which, together with the condition that $r_{b}$ has only real
values leads to \cite{Awad}: 
\begin{equation*}
n\leq \left( \frac{3-2\sqrt{2}}{30}\right) ^{\frac{1}{2}}l
\end{equation*}

Let us notice that we obtained two bolt solutions, corresponding to the
different signs in the expression of $r_b$.

\begin{itemize}
\item $U(1)$ fibration over $S^{2}\times T^{2}$

The metric is given by: 
\begin{eqnarray}
ds^{2} &=&-F(r)(dt-2n_{1}\cos \theta _{1}d\phi _{1}-2n_{2}\theta _{2}d\phi
_{2})^{2}+F^{-1}(r)dr^{2}  \notag \\
&&+(r^{2}+n_{1}^{2})(d\theta _{1}^{2}+\sin ^{2}\theta _{1}d\phi
_{1}^{2})+(r^{2}+n_{2}^{2})(d\theta _{2}^{2}+d\phi _{2}^{2})  \label{64}
\end{eqnarray}%
where: 
\begin{eqnarray}
F(r) &=&\frac{%
3r^{6}+(l^{2}+5n_{2}^{2}+10n_{1}^{2})r^{4}+3(n_{2}^{2}l^{2}+10n_{1}^{2}n_{2}^{2}+n_{1}^{2}l^{2}+5n_{1}^{4})r^{2}%
}{3(r^{2}+n_{1}^{2})(r^{2}+n_{2}^{2})l^{2}}  \notag \\
&&+\frac{6ml^{2}r-3n_{1}^{2}n_{2}^{2}(l^{2}+5n_{1}^{2})}{%
3(r^{2}+n_{1}^{2})(r^{2}+n_{2}^{2})l^{2}}  \label{65}
\end{eqnarray}%
Here the above metric is a solution of vacuum Einstein field equations with
cosmological constant if and only if: 
\begin{equation}
(n_{1}^{2}-n_{2}^{2})\lambda =2  \label{66}
\end{equation}%
Notice that if either the NUT charges are equal or if cosmological constant
vanishes, the above $U(1)$ fibration over $S^{2}\times T^{2}$ is not a
solution of the vacuum Einstein field equations.
\end{itemize}

In order to analyze the singularities of the above metrics let us consider
their Euclidean sections, obtained by analytic continuation of the
coordinate $t\rightarrow i\chi $ and of the parameters $n_{j}\rightarrow
in_{j}$ with $j=1,2$. If $\lambda =-10/l^{2}$ then the above constraint
equation becomes $n_{1}^{2}=n_{2}^{2}+\frac{5}{l^{2}}$. Then the NUT
solutions corresponds to a two-dimensional fixed-point set of $\frac{%
\partial }{\partial \chi }$ located at $r=n_{1}$. The periodicity of the
coordinate $\chi $ can be shown to be $8\pi n_{1}$ (ensuring the avoidance
of string singularities) while the value of the mass parameter is: 
\begin{equation*}
m=m_{b}=-\frac{n_{1}}{15l^{2}}(3l^{4}-40n_{1}^{2}l^{2}+120n_{1}^{4})
\end{equation*}%
For this value of the mass parameter the metric is well-behaved at $r=n_{1}$
as one can see by calculating some of the curvature invariants.

For the bolt solutions the fixed-point set is four-dimensional and it is
located at: 
\begin{equation*}
r=r_{b}=\frac{1}{20n}(l^{2}\pm \sqrt{l^{4}-80n_{1}^{2}l^{2}+400n_{1}^{4}})
\end{equation*}%
while the value of the mass parameter is given by: 
\begin{equation}
m=m_{b}=-\frac{%
15r_{b}^{6}+(10l^{2}-75n_{1}^{2})r_{b}^{4}+(3l^{4}-60n_{1}^{2}l^{2}+225n_{1}^{4})r_{b}^{2}+3n_{1}^{2}(l^{4}-10n_{1}^{2}l^{2}+25n_{1}^{4})%
}{30l^{2}r_{b}}
\end{equation}

Again, we have two kinds of bolt solutions given by the two roots $r=r_b$.

\begin{itemize}
\item $U(1)$ fibration over $S^{2}\times H^{2}$

The metric is given by: 
\begin{eqnarray}
ds^{2} &=&-F(r)(dt-2n_{1}\cos \theta _{1}d\phi _{1}-2n_{2}\cosh \theta
_{2}d\phi _{2})^{2}+F^{-1}(r)dr^{2}  \notag \\
&&+(r^{2}+n_{1}^{2})(d\theta _{1}^{2}+\sin ^{2}\theta _{1}d\phi
_{1}^{2})+(r^{2}+n_{2}^{2})(d\theta _{2}^{2}+\sinh ^{2}\theta _{2}d\phi
_{2}^{2})  \label{67}
\end{eqnarray}%
where: 
\begin{eqnarray}
F(r) &=&\frac{%
3r^{6}+(l^{2}+5n_{2}^{2}+10n_{1}^{2})r^{4}+3(n_{2}^{2}l^{2}+10n_{1}^{2}n_{2}^{2}+n_{1}^{2}l^{2}+5n_{1}^{4})r^{2}%
}{3(r^{2}+n_{1}^{2})(r^{2}+n_{2}^{2})l^{2}}  \notag \\
&&+\frac{6ml^{2}r-3n_{1}^{2}n_{2}^{2}(l^{2}+5n_{1}^{2})}{%
3(r^{2}+n_{1}^{2})(r^{2}+n_{2}^{2})l^{2}}  \label{68}
\end{eqnarray}%
Here the above metric is a solution of vacuum Einstein field equations with
cosmological constant if and only if: 
\begin{equation}
(n_{1}^{2}-n_{2}^{2})\lambda =4  \label{69}
\end{equation}%
Again, if the cosmological constant is zero or if the NUT charges are equal,
the above $U(1)$ fibration over $S^{2}\times H^{2}$ is not solution of
vacuum Einstein field equations.
\end{itemize}

As before we analytically continue $t\rightarrow i\chi $ and $%
n_{j}\rightarrow in_{j}$, where $j=1,2$. The constraint equation becomes in
this case $n_{1}^{2}=n_{2}^{2}+\frac{2l^{2}}{5}$. The NUT solution will
correspond to a two-dimensional fixed-point set of the vector $\frac{%
\partial }{\partial \chi }$, located at $r=n_{1}$. The value of the mass
parameter is given by: 
\begin{equation*}
m=m_{b}=-\frac{2n_{1}}{5l^{2}}(l^{4}-10n_{1}^{2}l^{2}+20n_{1}^{4})
\end{equation*}%
and the periodicity of the coordinate $\chi $ is $8\pi n_{1}$. Notice that
for this value of the mass parameter the metric will be well-behaved in the
vicinity of $r=n_{1}$.

The bolt solution will correspond to a four-dimensional fixed-point set
located at: 
\begin{equation*}
r=r_{b}=\frac{1}{20n}(l^{2}\pm \sqrt{l^{4}-80n_{1}^{2}l^{2}+400n_{1}^{4}})
\end{equation*}%
while the value of the mass parameter is given by: 
\begin{equation}
m=m_{b}=-\frac{%
15r_{b}^{6}+(10l^{2}-75n_{1}^{2})r_{b}^{4}+(3l^{4}-60n_{1}^{2}l^{2}+225n_{1}^{4})r_{b}^{2}+3n_{1}^{2}(l^{4}-10n_{1}^{2}l^{2}+25n_{1}^{4})%
}{30l^{2}r_{b}}
\end{equation}

Notice that we obtain two bolt solutions, that corresponds to the different
signs in the expression of $r=r_b$.

\begin{itemize}
\item $U(1)$ fibration over $T^{2}\times T^{2}$

The metric is given by: 
\begin{eqnarray}
ds^{2} &=&-F(r)(dt-2n_{1}\theta _{1}d\phi _{1}-2n_{2}\theta _{2}d\phi
_{2})^{2}+F^{-1}(r)dr^{2}  \notag \\
&&+(r^{2}+n_{1}^{2})(d\theta _{1}^{2}+d\phi
_{1}^{2})+(r^{2}+n_{2}^{2})(d\theta _{2}^{2}+d\phi _{2}^{2})  \label{610}
\end{eqnarray}%
where: 
\begin{equation}
F(r)=\frac{%
3r^{6}+5(n_{2}^{2}+2n_{1}^{2})r^{4}+15n_{1}^{2}(n_{1}^{2}+2n_{2}^{2})r^{2}+6ml^{2}r-15n_{1}^{4}n_{2}^{2}%
}{3(r^{2}+n_{1}^{2})(r^{2}+n_{2}^{2})l^{2}}  \label{611}
\end{equation}%
Here the above metric is a solution of vacuum Einstein field equations with
cosmological constant if and only if: 
\begin{equation}
(n_{1}^{2}-n_{2}^{2})\lambda =0  \label{612}
\end{equation}%
As in the case of the circle fibration over $S^{2}\times S^{2}$, if the
cosmological constant is zero, we can have two different parameters $n_{1}$
and $n_{2}$. If the cosmological constant is not zero then the above $U(1)$
fibration over $T^{2}\times T^{2}$ is a solution of the vacuum Einstein
field equations if and only if $n_{1}^{2}=n_{2}^{2}$.
\end{itemize}

If we consider the Euclidean section in this case, notice that there are no
Misner strings and that $F_{E}(r)$ becomes zero only if $r=0$. The solution
is singular at $r=n$, where $n$ is the greatest of the NUT charges $n_{1}$
and $n_{2}$.

Let us consider now the more interesting case in which the cosmological
constant is non-zero. Then the constraint equation imposes $n_{1}=n_{2}=n$.
There is a curvature singularity at $r=n$. However it can be readily checked
that if the mass parameter is given by $m=m_{b}=-\frac{8n^{5}}{l^{2}}$ then
the metric is well-behaved at $r=n$, and it corresponds to a NUT solution.
The bolt solutions have a four-dimensional fixed-point set located at $%
r_{b}>n$ and we find that in this case the mass parameter is given by: 
\begin{equation*}
m_{b}=-\frac{r_{b}^{6}-5n^{2}r_{b}^{4}+15n^{4}r_{b}^{2}+5n^{6}}{2l^{2}r_{b}}
\end{equation*}%
Regularity at the bolt requires the period of $\chi $ to be: 
\begin{equation*}
\beta =\frac{4\pi l^{2}r_{b}}{5(r_{b}^{2}-n^{2})}
\end{equation*}

\begin{itemize}
\item $U(1)$ fibration over $T^{2}\times H^{2}$

The metric is given by: 
\begin{eqnarray}
ds^{2} &=&-F(r)(dt-2n_{1}\theta _{1}d\phi _{1}-2n_{2}\cosh \theta _{2}d\phi
_{2})^{2}+F^{-1}(r)dr^{2}  \notag \\
&&+(r^{2}+n_{1}^{2})(d\theta _{1}^{2}+d\phi
_{1}^{2})+(r^{2}+n_{2}^{2})(d\theta _{2}^{2}+\sinh ^{2}\theta _{2}d\phi
_{2}^{2})  \label{613}
\end{eqnarray}%
where: 
\begin{equation}
F(r)=\frac{%
3r^{6}+5(n_{2}^{2}+2n_{1}^{2})r^{4}+15n_{1}^{2}(n_{1}^{2}+2n_{2}^{2})r^{2}+6ml^{2}r-15n_{1}^{4}n_{2}^{2}%
}{3(r^{2}+n_{1}^{2})(r^{2}+n_{2}^{2})l^{2}}  \label{614}
\end{equation}%
Here the above metric is a solution of vacuum Einstein field equations with
cosmological constant if and only if: 
\begin{equation}
(n_{1}^{2}-n_{2}^{2})\lambda =2  \label{615}
\end{equation}%
and so for vanishing cosmological constant or equal NUT charges this $U(1)$
fibration over $T^{2}\times H^{2}$ is not a solution of the vacuum Einstein
field equations.
\end{itemize}

To analyze the singularities of the above spacetimes we shall consider their
Euclidean sections, which are obtained by the following analytical
continuations $t\rightarrow i\chi $ and $n_{j}\rightarrow in_{j}$, with $%
j=1,2$. The constraint equation will become in this case $%
n_{1}^{2}=n_{2}^{2}+\frac{l^{2}}{5}$ and generically there is a curvature
singularity localized at $r=n_{1}$. However, if the value of the mass
parameter is given by: 
\begin{equation*}
m_{p}=-\frac{4n_{1}^{3}(6n_{1}^{2}-l^{2})}{3l^{2}}
\end{equation*}%
the the metric becomes well-behaved at $r=n_{1}$. This corresponds to the
NUT solution for which the fixed-point set of $\partial_\chi$ is
two-dimensional and located at $r=n_1$. Solutions with bolts correspond to
four-dimensional fixed-point sets located at $r_b>n_1$, for which the mass
parameter is given by: 
\begin{eqnarray}
m_b&=&\frac{-3r_b^6+(15n^2-l^2)r_b^4+3n^2(2l^2-15n^2)r_b^2+3n^4(l^2-5n^2)}{%
6l^2r_b}
\end{eqnarray}
and the periodicity of $\chi$ is: 
\begin{eqnarray}
\beta&=&\frac{4\pi l^2r_b}{5(r_b^2-n^2)}
\end{eqnarray}

\begin{itemize}
\item $U(1)$ fibration over $H^2\times H^2$

The metric is given by: 
\begin{eqnarray}
ds^{2} &=&-F(r)(dt-2n_{1}\cosh \theta _{1}d\phi _{1}-2n_{2}\cosh \theta
_{2}d\phi _{2})^{2}+F^{-1}(r)dr^{2}  \notag \\
&&+(r^{2}+n_{1}^{2})(d\theta _{1}^{2}+\sinh ^{2}\theta _{1}d\phi
_{1}^{2})+(r^{2}+n_{2}^{2})(d\theta _{2}^{2}+\sinh ^{2}\theta _{2}d\phi
_{2}^{2})  \label{616}
\end{eqnarray}%
where: 
\begin{eqnarray}
F(r) &=&\frac{%
3r^{6}+(5n_{2}^{2}+10n_{1}^{2}-l^{2})r^{4}+3(-n_{2}^{2}l^{2}+10n_{1}^{2}n_{2}^{2}-n_{1}^{2}l^{2}+5n_{1}^{4})r^{2}%
}{3(r^{2}+n_{1}^{2})(r^{2}+n_{2}^{2})l^{2}}  \notag \\
&&+\frac{6ml^{2}r+3n_{1}^{2}n_{2}^{2}(l^{2}-5n_{1}^{2})}{%
3(r^{2}+n_{1}^{2})(r^{2}+n_{2}^{2})l^{2}}  \label{617}
\end{eqnarray}%
Here the above metric is a solution of vacuum Einstein field equations with
cosmological constant ($\lambda =-\frac{10}{l^{2}}$) if and only if: 
\begin{equation}
(n_{1}^{2}-n_{2}^{2})\lambda =0  \label{618}
\end{equation}%
Hence either $n_{1}^{2}=n_{2}^{2}$ or the cosmological constant vanishes.
\end{itemize}

Consider now the Euclidean sections of the above spacetimes obtained by
analytical continuation of the coordinate $t\rightarrow i\chi $ and of the
parameters $n_{j}\rightarrow in_{j}$, with $j=1,2$. If the cosmological
constant is zero one can have different values for the NUT charge parameters 
$n_{1}$ and $n_{2}$. Let us assume that $n_{1}>n_{2}$. Then we obtain a NUT
solution for which the fixed-point set of the isometry $\frac{\partial }{%
\partial \chi }$ is two-dimensional and localized at $r=n_{1}$. In this case
the mass parameter is given by $m_{b}=-\frac{n_{1}(3n_{2}^{2}+n_{1}^{2})}{3}$
and the periodicity of the coordinate $\chi $ is given by $8\pi n_{1}$. One
can easily check that for this value of the mass parameter the metric is
well-behaved at $r=n_{1}$. The corresponding bolt solution has a
four-dimensional fixed-point set located at $r=r_{b}=\frac{2n_{1}}{k}$,
where $k$ is an integer number such that the periodicity of the coordinate $%
\chi $ is given by $\frac{8\pi n_{1}}{k}$. In order to avoid the curvature
singularity at $r=n_{1}$ we impose $r_{b}>n_{1}$ hence we must have $k=1$.
The mass parameter is then given by $m_{b}=\frac{%
n_{1}(4n_{1}^{2}-15n_{2}^{2})}{12}$.

If the cosmological constant is non-zero then the above constraint equation
will impose $n_{1}=n_{2}=n$. There is a NUT solution which corresponds to a
zero-dimensional fixed-point set of the $\frac{\partial }{\partial \chi }$
isometry, the NUT being located at $r=n$. The periodicity of the coordinate $%
\chi $ is $12\pi n$ and the mass parameter has the value $m=m_{b}=-\frac{%
4n^{3}(l^{2}+6n^{2})}{3l^{2}}$. There is a curvature singularity at $r=n$ as
can be checked by calculating some of the curvature invariants at $r=n$.

The bolt solution corresponds to a four-dimensional fixed-point set, which
is located at: 
\begin{eqnarray}
r=r_b&=&\frac{1}{30n}(-l^2+\sqrt{l^4+180n^2l^2+900n^4})
\end{eqnarray}
while the mass parameter is given by: 
\begin{eqnarray}
m_b&=&-\frac{3r_b^6-(15n^2+l^2)r_b^4+3n^2(2l^2+15n^2)r_b^2+3n^4(l^2+5n^2)}{%
6l^2r_b}
\end{eqnarray}

Finally, we can consider circle fibrations over $CP^{2}$. In this case the
metric is given by: 
\begin{equation}
ds^{2}=-F(r)(dt+A)^{2}+F^{-1}(r)dr^{2}+(r^{2}+n^{2})d\Sigma ^{2}  \label{645}
\end{equation}%
The explicit form of $A$ and $d\Sigma ^{2}$ in this case was given in \cite%
{Awad}: 
\begin{equation}
d\Sigma ^{2}=\frac{du^{2}}{\left( 1+\frac{u^{2}}{6}\right) ^{2}}+\frac{u^{2}%
}{4\left( 1+\frac{u^{2}}{6}\right) ^{2}}(d\psi +\cos \theta d\phi )^{2}+%
\frac{u^{2}}{4\left( 1+\frac{u^{2}}{6}\right) }(d\theta ^{2}+\sin ^{2}\theta
d\phi ^{2})  \label{cp1}
\end{equation}%
and 
\begin{equation}
A=\frac{u^{2}n}{2\left( 1+\frac{u^{2}}{6}\right) ^{2}}(d\psi +\cos \theta
d\phi )  \label{cp2}
\end{equation}%
while the expression for $F(r)$ is the same as in the $S^{2}\times S^{2}$
case. A singularity analysis of this metric was given in \cite{Awad}.

\subsection{Fibrations over `partial' base factors}

Another class of solutions is given for base spaces that are products of $2$%
-dimensional Einstein manifolds and the $U(1)$ fibration is taken over only
one of the components of the product.

\begin{itemize}
\item ($U(1)$ fibration over $S^2$)$\times S^2$

The metric is written in the form: 
\begin{eqnarray}
ds^{2} &=&-F(r)(dt-2n\cos \theta _{1}d\phi _{1})^{2}+F^{-1}(r)dr^{2}  \notag
\\
&&+(r^{2}+n^{2})(d\theta _{1}^{2}+\sin ^{2}\theta _{1}d\phi _{1}^{2})+\alpha
r^{2}(d\theta _{2}^{2}+\sin ^{2}\theta _{2}d\phi _{2}^{2})  \label{619}
\end{eqnarray}%
where to satisfy the field equations we must have 
\begin{equation}
\alpha =\frac{2}{2-\lambda n^{2}}  \label{620}
\end{equation}%
\begin{equation}
F(r)=\frac{3r^{5}+(l^{2}+10n^{2})r^{3}+3n^{2}(l^{2}+5n^{2})r-6ml^{2}}{%
3rl^{2}(r^{2}+n^{2})}  \label{621}
\end{equation}%
The metric (\ref{619}) is a solution of the vacuum Einstein field equations
for any values of $n$ or $\lambda=-\frac{10}{l^2}$. However, in order retain
a Lorentzian metric signature we must ensure that $\alpha >0$, which
translates in our case $\lambda n^{2}<2$. We have given above the form of $%
F(r)$ using a negative cosmological constant and in this case this condition
is superfluous; however we can also use a positive cosmological constant (we
have to analytically continue $l\rightarrow il$ in $F(r)$) as long as the
above condition is satisfied. The Euclidian section is obtained by taking
the analytical continuation $t\rightarrow i\chi$ and $n\rightarrow in$. The
NUT solution corresponds to a two-dimensional fixed-point set of the vector $%
\frac{\partial}{\partial\chi}$ located at $r=n$. The periodicity of the $%
\chi $ coordinate is in this case equal to $8\pi n$ and the value of the
mass parameter is fixed to $m_b=\frac{n^3(l^2-4n^2)}{3l^2}$. Furthermore, if 
$n=\frac{l}{2}$ then curvature singularity present at $r=n$ disappears and
we obtain the $AdS$ spacetime. The bolt spacetime has a four-dimensional
fixed-point set of $\frac{\partial}{\partial\chi}$ located at $r=r_b$ with: 
\begin{eqnarray}
r_b&=&\frac{l^2\pm\sqrt{k^2l^4-80n^2l^2+400n^4}}{20n}
\end{eqnarray}
while the value of the mass parameter is: 
\begin{eqnarray}
m_b&=&-\frac{3r_b^5+(l^2-10n^2)r_b^3-3n^2(l^2-5n^2)r_b}{6l^2}
\end{eqnarray}
where the periodicity of the coordinate $\chi$ is $\frac{8\pi n}{k}$, where $%
k$ is an integer.

\item ($U(1)$ fibration over $S^2$)$\times T^2$

In this case we have 
\begin{eqnarray}
ds^{2} &=&-F(r)(dt-2n\cos \theta _{1}d\phi _{1})^{2}+F^{-1}(r)dr^{2}  \notag
\\
&&+(r^{2}+n^{2})(d\theta _{1}^{2}+\sin ^{2}\theta _{1}d\phi _{1}^{2})+\alpha
r^{2}(d\theta _{2}^{2}+d\phi _{2}^{2})  \label{622}
\end{eqnarray}%
where 
\begin{equation}
F(r)=\frac{-3r^{5}+(l^{2}-10n^{2})r^{3}+3n^{2}(l^{2}-5n^{2})r+6ml^{2}}{%
3rl^{2}(r^{2}+n^{2})}  \label{623}
\end{equation}%
and the vacuum Einstein field equations with cosmological constant are
satisfied if and only if: 
\begin{equation}
\alpha (2-\lambda n^{2})=0  \label{624}
\end{equation}%
Since $\alpha $ cannot be zero we must restrict the values of $n$ and $%
\lambda $ such that $\lambda n^{2}=2$, forcing a positive cosmological
constant. The Euclidian section is obtained by taking the analytical
continuation $t\rightarrow i\chi$ and $n\rightarrow in$ and $l\rightarrow il$
with $l=\sqrt{5}n$. Notice that in this case the NUT solution corresponds to
a two-dimensional fixed-point set of the vector field $\frac{\partial}{%
\partial\chi}$ located at $r=n$. The periodicity of the $\chi$ coordinate is
in this case equal to $8\pi n$ and the value of the mass parameter is fixed
to $m_b=\frac{n^3}{15}$. For this value of the mass parameter the solution
is regular at the nut location. The bolt spacetime has a four-dimensional
fixed-point set of $\frac{\partial}{\partial\chi}$ located at $r_b=\frac{kn}{%
2}$ and the value of the mass parameter is $m_b=\frac{k^3n^3(20-3k^2)}{960}$
where the periodicity of the coordinate $\chi$ is $\frac{8\pi n}{k}$, where $%
k$ is an integer. To ensure that $r_b>n$ we have to take $k>3$ and in this
way the curvature singularity at $r=n$ is avoided as well.

\item ($U(1)$ fibration over $S^{2}$)$\times H^{2}$

The metric is written in the form: 
\begin{eqnarray}
ds^{2} &=&-F(r)(dt-2n\cos \theta _{1}d\phi _{1})^{2}+F^{-1}(r)dr^{2}  \notag
\\
&&+(r^{2}+n^{2})(d\theta _{1}^{2}+\sin ^{2}\theta _{1}d\phi _{1}^{2})+\alpha
r^{2}(d\theta _{2}^{2}+\sinh ^{2}\theta _{2}d\phi _{2}^{2})  \label{625}
\end{eqnarray}%
where 
\begin{equation}
\alpha =\frac{2}{\lambda n^{2}-2}  \label{626}
\end{equation}%
\begin{equation}
F(r)=\frac{-3r^{5}+(l^{2}-10n^{2})r^{3}+3n^{2}(l^{2}-5n^{2})r+6ml^{2}}{%
3rl^{2}(r^{2}+n^{2})}  \label{626b}
\end{equation}%
are necessary for the vacuum Einstein field equations to hold. The condition 
$\alpha >0$ implies that $\lambda n^{2}>2$, again yielding a positive
cosmological constant. The Euclidian section is obtained by taking the
analytical continuation $t\rightarrow i\chi$ and $n\rightarrow in$ while in
order to keep $\alpha>0$ we must continue $l\rightarrow il$ . The NUT
solution corresponds to a two-dimensional fixed-point set of the vector $%
\frac{\partial}{\partial\chi}$ located at $r=n$. The periodicity of the $%
\chi $ coordinate is in this case equal to $8\pi n$ and the value of the
mass parameter is fixed to $m_b=\frac{n^3(l^2-4n^2)}{3l^2}$ . Furthermore,
if $n=\frac{l}{2}$ then curvature singularity present at $r=n$ disappears
and we obtain the six-dimensional $AdS$ spacetime. The bolt spacetime has a
four-dimensional fixed-point set of $\frac{\partial}{\partial\chi}$ located
at $r=r_b$ with: 
\begin{eqnarray}
r_b&=&\frac{kl^2\pm\sqrt{k^2l^4-80n^2l^2+400n^4}}{20n}
\end{eqnarray}
while the value of the mass parameter is: 
\begin{eqnarray}
m_b&=&-\frac{3r_b^5+(l^2-10n^2)r_b^3-3n^2(l^2-5n^2)r_b}{6l^2}
\end{eqnarray}
where the periodicity of the coordinate $\chi$ is $\frac{8\pi n}{k}$, where $%
k$ is an integer.

\item ($U(1)$ fibration over $T^2$)$\times S^2$

The metric is written in the form: 
\begin{eqnarray}
ds^{2} &=&-F(r)(dt-2n\theta _{1}d\phi _{1})^{2}+F^{-1}(r)dr^{2}  \notag \\
&&+(r^{2}+n^{2})(d\theta _{1}^{2}+d\phi _{1}^{2})+\alpha r^{2}(d\theta
_{2}^{2}+\sin ^{2}\theta _{2}d\phi _{2}^{2})  \label{628}
\end{eqnarray}%
where now 
\begin{equation}
\alpha =-\frac{2}{\lambda n^{2}}  \label{629}
\end{equation}%
\begin{equation}
F(r)=\frac{3r^{5}+10n^{2}r^{3}+15n^{4}r-6ml^{2}}{3rl^{2}(r^{2}+n^{2})}
\label{630}
\end{equation}%
While the above metric is a solution of the vacuum Einstein field equations
for any values of $n$ or $\lambda $, the condition $\alpha >0$ implies that $%
\lambda <0$.

The Euclidian section of this solution is obtained by analytical
continuations $t\rightarrow i\chi $, $n\rightarrow in$ and $l\rightarrow il$
(in order to preserve $\alpha >0$). The circle fibration is trivial: in this
case we obtain two-dimensional fixed point sets of $\partial _{\chi }$ (or
`nuts') for $m_{n}=\frac{4n^{5}}{3l^{2}}$. The bolt solutions have a
four-dimensional fixed-point set located at $r_{b}>n$ and we find that in
this case the mass parameter is given by: 
\begin{equation*}
m_{b}=\frac{3r_{b}^{5}-10n^{2}r_{b}^{3}+15n^{4}r_{b}}{6l^{2}}
\end{equation*}%
The Euclidian regularity at the bolt requires the period of $\chi $ to be: 
\begin{equation*}
\beta =\frac{4\pi l^{2}r_{b}}{5(r_{b}^{2}-n^{2})}
\end{equation*}

\item ($U(1)$ fibration over $T^2$)$\times T^2$

The metric is written in the form: 
\begin{eqnarray}
ds^{2} &=&-F(r)(dt-2n\theta _{1}d\phi _{1})^{2}+F^{-1}(r)dr^{2}  \notag \\
&&+(r^{2}+n^{2})(d\theta _{1}^{2}+d\phi _{1}^{2})+r^{2}(d\theta
_{2}^{2}+d\phi _{2}^{2})  \label{631}
\end{eqnarray}%
where 
\begin{equation}
F(r)=\frac{3r^{5}+10n^{2}r^{3}+15n^{4}r-6ml^{2}}{3rl^{2}(r^{2}+n^{2})}
\label{632}
\end{equation}%
and the vacuum Einstein field equations force the condition $\lambda n^{2}=0$
where $\lambda =\pm \frac{10}{l^{2}}$. Hence a $U(1)$ fibration over $T^{2}$
exists if and only if $\lambda=0$.

The Euclidian section of this solution is obtained by analytical
continuations $t\rightarrow i\chi$, $n\rightarrow in$. Since $F_E(r)=\frac{2m%
}{(r^2-n^2)r}$ we notice that $F_E(r)$ cannot be zero for any value of the
radius $r$. Generically there is a curvature singularity located at $r=n$ at
which point the fixed-point set of $\partial_\chi$ is two-dimensional. The
`bolt' solutions correspond to four-dimensional fixed-point sets located at $%
r_b>n$ (which also avoids the curvature singularity). In both cases there
are no restriction of the periodicity of $\chi$.

\item ($U(1)$ fibration over $T^2$)$\times H^2$

The metric is written in the form: 
\begin{eqnarray}
ds^{2} &=&-F(r)(dt-2n\theta _{1}d\phi _{1})^{2}+F^{-1}(r)dr^{2}  \notag \\
&&+(r^{2}+n^{2})(d\theta _{1}^{2}+d\phi _{1}^{2})+\alpha r^{2}(d\theta
_{2}^{2}+\sinh ^{2}\theta _{2}d\phi _{2}^{2})  \label{633}
\end{eqnarray}%
where in this case%
\begin{equation}
\alpha =\frac{2}{\lambda n^{2}}  \label{634}
\end{equation}%
\begin{equation}
F(r)=\frac{-3r^{5}-10n^{2}r^{3}-15n^{4}r+6ml^{2}}{3rl^{2}(r^{2}+n^{2})}
\label{635}
\end{equation}%
The above metric is a solution of vacuum Einstein field equations for any
values of $n$ or $\lambda $. Again the condition $\alpha >0$ implies $%
\lambda >0$.

The Euclidian section of this solution is obtained by analytical
continuations $t\rightarrow i\chi$, $n\rightarrow in$ and $l\rightarrow il$
(in order to preserve $\alpha>0$). The circle fibration is trivial in this
case we obtain two-dimensional fixed point sets of $\partial_\chi$ (or
`nuts') for $m_n=-\frac{4n^5}{3l^2}$. The bolt solutions have a
four-dimensional fixed-point set located at $r_b>n$ and we find that in this
case the mass parameter is given by: 
\begin{eqnarray}
m_b&=&-\frac{3r_b^5-10n^2r_b^3+15n^4r_b}{6l^2}
\end{eqnarray}
The Euclidian regularity at the bolt requires the period of $\chi$ to be: 
\begin{eqnarray}
\beta&=&\frac{4\pi l^2r_b}{5(r_b^2-n^2)}
\end{eqnarray}

\item ($U(1)$ fibration over $H^2$)$\times S^2$

The metric is written in the form: 
\begin{eqnarray}
ds^{2} &=&-F(r)(dt-2n\cosh \theta _{1}d\phi _{1})^{2}+F^{-1}(r)dr^{2}  \notag
\\
&&+(r^{2}+n^{2})(d\theta _{1}^{2}+\sinh ^{2}\theta _{1}d\phi
_{1}^{2})+\alpha r^{2}(d\theta _{2}^{2}+\sin ^{2}\theta _{2}d\phi _{2}^{2})
\label{636}
\end{eqnarray}%
where 
\begin{equation}
\alpha =-\frac{2}{2+\lambda n^{2}}  \label{637}
\end{equation}%
\begin{equation}
F(r)=\frac{3r^{5}+(-l^{2}+10n^{2})r^{3}+3n^{2}(-l^{2}+5n^{2})r-6ml^{2}}{%
3rl^{2}(r^{2}+n^{2})}  \label{638}
\end{equation}%
The above metric is a solution of vacuum Einstein field equations for any
values of $n$ or $\lambda$. Now the condition $\alpha >0$ is equivalent to $%
\lambda n^{2}<-2$ and so we must have a negative cosmological constant.

The Euclidian sections of this solution correspond to the analytical
continuations $t\rightarrow i\chi$ and $n\rightarrow in$. However, in order
to keep the sign of $\alpha$ positive we have to analytically continue $%
l\rightarrow il$ and impose the constraint relation $5n^2>l^2$. The NUT
solution has a two-dimensional fixed-point set of $\partial_\chi$ (a `nut')
located at $r=n$ and the value of the mass parameter is $m_n=\frac{%
n^3(4n^2-l^2)}{3l^2}$, the periodicity of $\chi$ being in this case $8\pi n$%
. For this value of the mass parameter the spacetime geometry is smooth at $%
r=n$. The bolt solutions have a four-dimensional fixed-point set of $%
\partial_\chi$ located at $r_b>n$ where: 
\begin{eqnarray}
r_b&=&\frac{kl^2\pm\sqrt{k^2l^4-80n^2l^2+400n^4}}{20n}
\end{eqnarray}
the mass parameter being given by: 
\begin{eqnarray}
m_b&=&\frac{3r_b^5+(l^2-10n^2)r_b^3-3n^2(l^2-5n^2)r_b}{6l^2}
\end{eqnarray}
while the periodicity of $\chi$ is given by $\frac{8\pi n}{k}$, $k$ being an
integer. Notice that we have obtained two bolt solutions each corresponding
to the different signs in $r_b$'s expression.

\item ($U(1)$ fibration over $H^2$)$\times T^2$

The metric is written in the form: 
\begin{eqnarray}
ds^{2} &=&-F(r)(dt-2n\cosh \theta _{1}d\phi _{1})^{2}+F^{-1}(r)dr^{2}  \notag
\\
&&+(r^{2}+n^{2})(d\theta _{1}^{2}+\sinh ^{2}\theta _{1}d\phi _{1}^{2})+
r^{2}(d\theta _{2}^{2}+d\phi _{2}^{2})  \label{639}
\end{eqnarray}%
where: 
\begin{equation}
F(r)=\frac{3r^{5}+(-l^{2}+10n^{2})r^{3}+3n^{2}(-l^{2}+5n^{2})r-6ml^{2}}{%
3rl^{2}(r^{2}+n^{2})}  \label{640}
\end{equation}%
The above metric is a solution of vacuum Einstein field equations with a
cosmological constant if and only if 
\begin{equation}
2+\lambda n^{2}=0  \label{641}
\end{equation}%
Hence we have to restrict the values of $n$ or $\lambda $ such that $\lambda
n^{2}=-2$ (that is $\lambda <0$).

The Euclidian section of this solution is obtained by analytical
continuation $t\rightarrow i\chi$, $n\rightarrow in$ and $l\rightarrow il$
such that $5n^2=l^2$. The NUT solution has a two-dimensional fixed-point set
of $\partial_\chi$ located at $r=n$, the mass parameter is given by $%
m_n=-n^3/15$ and the periodicity of $\chi$ is $8\pi n$. Notice that for this
value of the mass parameter the geometry is smooth at the nut's location.
The bolt solution has a four-dimensional fixed-point set at $r_b=\frac{kn}{2}
$, the mass parameter is given by: 
\begin{eqnarray}
m_b&=&\frac{k^3n^3(3k^2-20)}{960}
\end{eqnarray}
the periodicity of $\chi$ being $8\pi n/k$. Since $r_b>n$ we must restrict
the values of the integer $k$ such that $k\geq 3$.

\item ($U(1)$ fibration over $H^2$)$\times H^2$

The metric is written in the form: 
\begin{eqnarray}
ds^{2} &=&-F(r)(dt-2n\cosh \theta _{1}d\phi _{1})^{2}+F^{-1}(r)dr^{2}  \notag
\\
&&+(r^{2}+n^{2})(d\theta _{1}^{2}+\sinh ^{2}\theta _{1}d\phi
_{1}^{2})+\alpha r^{2}(d\theta _{2}^{2}+\sinh ^{2}\theta _{2}d\phi _{2}^{2})
\label{642}
\end{eqnarray}%
where the vacuum Einstein field equations imply 
\begin{equation}
\alpha =\frac{2}{\lambda n^{2}+2}  \label{643}
\end{equation}%
\begin{equation}
F(r)=\frac{3r^{5}+(-l^{2}+10n^{2})r^{3}+3n^{2}(-l^{2}+5n^{2})r-6ml^{2}}{%
3rl^{2}(r^{2}+n^{2})}  \label{644}
\end{equation}%
for any values of $n$ or $\lambda $. In this case the restriction $\alpha >0$
leads to $\lambda n^{2}>-2$. We can have a negative cosmological constant as
long as this relation is satisfied; alternatively we can have a positive
cosmological constant (by analytically continuing $l\rightarrow il$ in the
above expression for $F(r)$).
\end{itemize}

The Euclidian section of this solution is obtained by analytical
continuation $t\rightarrow i\chi$ and $n\rightarrow in$. We can choose to
analytically continue the parameter $l$ to obtain a solution with positive
cosmological constant, however in this case we must impose the restriction $%
\alpha>0$ which amounts to $5n^2<l^2$. Let us consider first the solution
with negative cosmological constant (that is, we do not analytically
continue $l$ in the initial solution). If we try to make $r=n$ into a fixed
point of $\partial_\chi$, we find that $F_E(r)$ becomes negative for $r$
close enough to (and bigger than) $n$. This means that $F_E(r)$ vanishes at
some value $r>n$ and instead of a nut we find a bolt. It is a situation
analogous with the four-dimensional case, where it was found in \cite%
{Chamblin} that there are no hyperbolic nuts. The bolt solution corresponds
to a four-dimensional fixed-point set of $\partial_\chi$ located at $r_b>n$.
The value of the mass parameter is given by: 
\begin{eqnarray}
m_b&=&\frac{3r_b^5-(l^2+10n^2)r_b^3+3n^2(l^2+5n^2)r_b}{6l^2}
\end{eqnarray}
where 
\begin{eqnarray}
r_b&=&\frac{-l^2+\sqrt{l^4+80n^2l^2+400n^4}}{20n}
\end{eqnarray}
while the periodicity of $\chi$ is $8\pi n$.

Let us consider now the Euclidian section obtained by further analytical
continuation $l\rightarrow il$. The NUT solution corresponds to a
two-dimensional fixed-point set of $\partial_\chi$ located at $r=n$ and the
value of the mass parameter is given by $m_n=\frac{n^3(l^2-4n^3)}{3l^2}$.
For this value of the mass parameter the spacetime geometry at the nut
location is smooth. The bolt solution corresponds to a four-dimensional
fixed-point set of $\partial_\chi$ located at: 
\begin{eqnarray}
r_b&=&\frac{l^2+\sqrt{l^4-80n^2l^2+400n^4}}{20n}
\end{eqnarray}
the mass parameter being given by: 
\begin{eqnarray}
m_b&=&-\frac{3r_b^5+(l^2-10n^2)r_b^3-3n^2(l^2-5n^2)r_b}{6l^2}
\end{eqnarray}
and the periodicity of $\chi$ is $8\pi n$. If we require $r_b>n$ then: 
\begin{eqnarray}
n&\leq&\left(\frac{1}{10}-\frac{\sqrt{3}}{20}\right)^{\frac{1}{2}}l .
\end{eqnarray}

\section{Taub-Nut-dS/AdS spacetimes in $7$ dimensions}

In seven dimensions the base space is five dimensional. We shall factorize
it in the form $B=M\times Y$, where $M$ is an even-dimensional Einstein
space endowed with an Einstein-K$\ddot{a}$hler metric. We have now two
possibilities: the factor $M$ can be two-dimensional or four-dimensional. If 
$M$ is two-dimensional we have three cases, which we can take to be a
sphere, a torus or a hyperboloid. In the four dimensional case we can use
products $M_{1}\times M_{2}$ of two-dimensional spaces or we can use $%
M=CP^{2}$. Let us consider next all these cases.

If $M$ is two dimensional we can take it to be a sphere, a torus or a
hyperboloid. Then the base factor can be factorized in the form $B=M\times Y$
with $Y$ a three dimensional riemannian space, which we shall take for
simplicity to be an Einstein space. For each choice of the factor $M$ we can
take $Y$ to be a sphere $S^{3}$, a torus $T^{3}$ or a hyperboloid $H^{3}$.
The ansatz for the Taub-NUT metric that we shall determine is given by: 
\begin{equation}
ds^{2}=-F(r)(dt+A)^{2}+F^{-1}(r)dr^{2}+(r^{2}+n^{2})g_{M}+\alpha r^{2}g_{Y}
\label{71}
\end{equation}%
where $g_{M}$ is the metric on the two-dimensional space $M$ and $g_{Y}$ is
a riemannian metric on the three dimensional space $Y$. We shall consider
the following cases:

\begin{itemize}
\item $M=S^{2}$ then $Y$ can be $S^{3}$, $T^{3}$ or $H^{3}$. The metric is
given by: 
\begin{eqnarray}
ds^{2} &=&-F(r)(dt^{2}+2n\cos \theta d\phi
)^{2}+F^{-1}(r)dr^{2}+(r^{2}+n^{2})(d\theta ^{2}+\sin ^{2}\theta d\phi ^{2})
\notag \\
&&+\alpha r^{2}g_{Y}  \label{72}
\end{eqnarray}%
where: 
\begin{equation}
F(r)=\frac{%
4r^{6}+(l^{2}+12n^{2})r^{4}+2n^{2}(l^{2}+6n^{2})r^{2}+4ml^{2}+n^{4}(l^{2}+6n^{2})%
}{4l^{2}r^{2}(r^{2}+n^{2})}  \label{73}
\end{equation}%
The cosmological constant is given by $\lambda =-\frac{15}{l^{2}}$\ and the
values of the parameters $\alpha $, $n$\ and $\lambda $\ are constrained via
the relation as follows: 
\begin{equation}
\alpha (5-2\lambda n^{2})=10s  \label{74}
\end{equation}%
where $s=1,0, -1,$ for $S^{3}$, $T^{3}$\ and $H^{3}$\ respectively. We must
have $\alpha >0$\ , which in turn imposes a joint constraint on $\lambda
n^{2} $, which can be satisfied in various ways depending on the value of $s$%
. Solutions for a positive value of $\lambda $\ can be obtained by
analytically continuing $l\rightarrow il$\ in the expression (\ref{73}) for
the function $F(r)$.

The Euclidian sections of the above solutions is obtained by the analytical
continuations $t\rightarrow i\chi$ and $n\rightarrow in$. We are free to
make the continuation $l\rightarrow il$ as long as the condition $\alpha>0$
is always satisfied. Since the singularity analysis of the metric does not
depend on this positivity condition we shall consider the two cases: in the
first case we analytically continue $n$ in the above expression of $F(r)$
(which corresponds to $\lambda<0$) and in the second case we perform the
analytical continuation of $n$ and $l$ (which corresponds to a positive
cosmological).

If the cosmological constant is negative $\lambda=-\frac{15}{l^2}$ then the
NUT solution corresponds to a three-dimensional fixed-point set of $%
\partial_\chi$ located at $r=n$, for which the value of the mass parameter
is $m_n=\frac{n^6}{2l^2}$. The bolt solutions correspond to a
five-dimensional fixed-point set of $\partial_\chi$ located at $r=r_b$
where: 
\begin{eqnarray}
r_b&=&\frac{l^2\pm\sqrt{l^4-96n^2l^2+576n^4}}{24n}
\end{eqnarray}
the value of the mass parameter is 
\begin{eqnarray}
m_b&=&-\frac{4r_b^6+(l^2-12n^2)r_b^4+2n^2(6n^2-l^2)r_b^2+n^4(l^2-6n^2)}{4l^2}
\end{eqnarray}
the periodicity of $\chi$ being $8\pi n$. In order to have $r_b>n$ we have
the condition: 
\begin{eqnarray}
n&\leq&\left(\frac{1}{12}-\frac{\sqrt{3}}{24}\right)^{\frac{1}{2}}l .
\end{eqnarray}

Consider now the case in which we make the further analytical continuation $%
l\rightarrow il$ (we obtain a solution with positive cosmological constant).
If we try to make $r=n$ into a fixed point of $\partial_\chi$ we find that $%
F_E(r)$ becomes positive for $r$ close enough to (and bigger than) $n$. This
means that $F_E(r)$ vanishes at some value $r>n$ and instead of a nut we
find a bolt. Hence in this case there are no nut solutions. The bolt
solutions correspond to a five-dimensional fixed-point set of $\partial_\chi$
located at $r=r_b>n$. Their mass parameter is given by: 
\begin{eqnarray}
m_b&=&\frac{4r_b^6-(l^2+12n^2)r_b^4+2n^2(l^2-6n^2)r_b^2-n^4(l^2-6n^2)}{4l^2}
\end{eqnarray}
the periodicity of $\chi$ being $\frac{4\pi l^2r_b}{l^2-6(r_b^2-n^2)} $.

\item $M=T^{2}$ then $Y$ can be $S^{3}$, $T^{3}$ or $H^{3}$. Then the metric
is given by: 
\begin{equation}
ds^{2}=-F(r)(dt^{2}+2n\theta d\phi
)^{2}+F^{-1}(r)dr^{2}+(r^{2}+n^{2})(d\theta ^{2}+d\phi ^{2})+\alpha
r^{2}g_{Y}  \label{75}
\end{equation}%
where (with $\lambda =-\frac{15}{l^{2}}$) we have%
\begin{equation}
F(r)=\frac{2r^{6}+6n^{2}r^{4}+6n^{4}r^{2}+4ml^{2}+3n^{6}}{%
2l^{2}r^{2}(r^{2}+n^{2})}  \label{76}
\end{equation}%
along with the constraint 
\begin{equation}
\alpha \lambda n^{2}=-5s  \label{77}
\end{equation}%
where $s=1,0, -1,$ for $S^{3}$, $T^{3}$\ and $H^{3}$\ respectively as
before. \ Note that if $Y=H^{3}$\ we must analytically continue $%
l\rightarrow il$\ in the expression for $F(r)$\ so as to have a positive
cosmological constant.

If $s=1$ then initially we have a negative cosmological constant and when we
go to the Euclidian section (by means of the analytical continuations $%
t\rightarrow i\chi$, $n\rightarrow in$)we must continue as well $%
l\rightarrow il$ to keep $\alpha>0$. Then the NUT solution corresponds to a
three-dimensional fixed-point set of $\partial_\chi$ located at $r=n$ with
the mass parameter $m_n=-\frac{n^6}{4l^2}$. The bolt solutions correspond to
a five-dimensional fixed-point set of $\partial_\chi$ located at $r=r_b>n$
with the mass parameter 
\begin{eqnarray}
m_b&=&\frac{2r_b^6-6n^2r_b^4+6n^4r_b^2-3n^6}{4l^2}
\end{eqnarray}
while the periodicity of $\chi$ is given by $\frac{2\pi l^2r_b}{3(r_b^2-n^2)}
$.

If $s=0$ then the cosmological constant must vanish. Notice that in this
case $F_E(r)$ does not vanish anywhere and that the solution is singular at $%
r=0$.

If $s=-1$ we obtain the Euclidian section by means of analytical
continuations $l\rightarrow i\chi$, $n\rightarrow in$ and $l\rightarrow il$
(preserving the condition $\alpha>0$). We obtain a NUT solution with a
three-dimensional fixed-point set of $\partial_\chi$ located at $r=n$ with
the mass parameter $m_n=\frac{n^6}{4l^2}$. Notice that for this value of the
mass parameter there is no curvature singularity located at $r=n$. We also
have bolt solutions with a five-dimensional fixed-point set of $%
\partial_\chi $ located at $r=r_b>n$ with the mass parameter 
\begin{eqnarray}
m_b&=&-\frac{2r_b^6-6n^2r_b^4+6n^4r_b^2-3n^6}{4l^2}
\end{eqnarray}
while the periodicity of $\chi$ is given by $\frac{2\pi l^2r_b}{3(r_b^2-n^2)}
$.

\item $M=H^{2}$ then $Y$ can be $S^{3}$, $T^{3}$ or $H^{3}$. Then the metric
is given by: 
\begin{equation}
ds^{2}=-F(r)(dt^{2}+2n\cosh \theta d\phi
)^{2}+F^{-1}(r)dr^{2}+(r^{2}+n^{2})(d\theta ^{2}+\sinh ^{2}\theta d\phi
^{2})+\alpha r^{2}g_{Y}  \label{78}
\end{equation}%
where as before $\lambda =-\frac{15}{l^{2}}$, 
\begin{equation}
F(r)=\frac{%
4r^{6}+(-l^{2}+12n^{2})r^{4}+2n^{2}(-l^{2}+6n^{2})r^{2}+4ml^{2}+n^{4}(-l^{2}+6n^{2})%
}{4l^{2}r^{2}(r^{2}+n^{2})}  \label{79}
\end{equation}%
and the values of the parameters $\alpha $, $n$ and $\lambda $ are
constrained via as follows: 
\begin{equation}
\alpha \left( 5+2\lambda n^{2}\right) =-10s  \label{710}
\end{equation}%
where $s=1,0, -1,$ for $S^{3}$, $T^{3}$\ and $H^{3}$\ respectively as
before; should a positive value of $\lambda $ be required we must
analytically continue $l\rightarrow il$\ in the above expression (\ref{79})
for $F(r)$.
\end{itemize}

If $s=1$ we obtain the Euclidian section by making the analytical
continuations $t\rightarrow i\chi$, $n\rightarrow in$ and $l\rightarrow il$
(again preserving the positivity of $\alpha$). The NUT solution corresponds
to a three-dimensional fixed-point set of $\partial_\chi$ located at $r=n$
with the mass parameter $m_n=-\frac{n^6}{2l^2}$. The periodicity of $\chi$
is $8\pi n$ and there is no curvature singularity at $r=n$. The bolt
solutions correspond to a five-dimensional fixed-point set of $\partial_\chi$
located at $r=r_b$ with the mass parameter 
\begin{eqnarray}
m_b&=&\frac{4r_b^6+(l^2-12n^2)r_b^4-2n^2(l^2-6n^2)r_b^2+n^4(l^2-6n^2)}{4l^2}
\end{eqnarray}
and 
\begin{eqnarray}
r_b&=&\frac{l\pm\sqrt{l^4-96n^2l^2+576n^4}}{24n}
\end{eqnarray}
while the periodicity of $\chi$ is $8\pi n$. If we require $r_b>n$ we obtain 
\begin{eqnarray}
n&<&\left(\frac{1}{12}-\frac{\sqrt{3}}{24}\right)^{\frac{1}{2}}l
\end{eqnarray}

If $s=0$ we have the constraint $l=\sqrt{6}n$ and to obtain the Euclidian
section we have make the analytical continuations $t\rightarrow i\chi$, $%
n\rightarrow in$. If we try to make $r=n$ into a fixed-point of $%
\partial_\chi$ we find that $F_E(r)$ becomes negative for $r$ close enough
to (and bigger than $n$), which means that $F_E(r)$ vanishes at some value $%
r>n$ and instead of having a nut we find a bolt. The bolt solutions
correspond to a five-dimensional fixed-point set of $\partial_\chi$ located
at $r=r_b$ while the mass parameter is: 
\begin{eqnarray}
m_b&=&-\frac{2r_b^6-9n^2r_b^4+12n^4r_b^2-6n^6}{12n^2}
\end{eqnarray}
and the periodicity of $\chi$ is $\frac{4\pi n^2r_b}{r_b^2-2n^2}$.

If $s=-1$ then we can have two cases, corresponding to a positive or a
negative cosmological constant. If we analytically continue $t\rightarrow
i\chi $ and $n\rightarrow in$ we obtain an Euclidian section with negative
cosmological constant; however $\alpha <0$ and so the solution is
unphysical. If we analytically continue $t\rightarrow i\chi $, $n\rightarrow
in$ and $l\rightarrow il$ we obtain a Euclidian section with positive
cosmological constant. In order to ensure the positivity of $\alpha $ we
must require that $6n^{2}>l^{2}$. The NUT solution corresponds to a three
dimensional fixed-point set of $\partial _{\chi }$ located at $r=n$, the
mass parameter is $m_{n}=-\frac{n^{6}}{2l^{2}}$ and the periodicity of $\chi 
$ is $8\pi n$. The bolt solutions have a five-dimensional fixed-point set of 
$\partial _{\chi }$ located at $r=r_{b}$, where: 
\begin{equation*}
r_{b}=\frac{kl\pm \sqrt{k^{2}l^{4}-96n^{2}l^{2}+576n^{4}}}{24n}
\end{equation*}%
the mass parameter is given by: 
\begin{equation*}
m_{b}=\frac{%
4r_{b}^{6}+(l^{2}-12n^{2})r_{b}^{4}-2n^{2}(l^{2}-6n^{2})r_{b}^{2}+n^{4}(l^{2}-6n^{2})%
}{4l^{2}}
\end{equation*}%
and the periodicity of $\chi $ is $\frac{8\pi n}{k}$, $k$ being an integer.

The second possibility in the ansatz (\ref{71})is to take the space $M$ four
dimensional and the space $Y$ one-dimensional. In this case we can further
factorize $M$ into products of two-dimensional spaces with constant
curvature $M=M_1\times M_2$, or we can consider $M=CP^2$.

Let us begin with the second case, namely $M=CP^{2}$. The metric is%
\begin{equation}
ds^{2}=-F(r)(dt+A)^{2}+F^{-1}(r)dr^{2}+(r^{2}+n^{2})d\Sigma ^{2}+r^{2}dy^{2}
\label{711}
\end{equation}%
where $A$ and $d\Sigma ^{2}$ are given in (\ref{cp1}) and (\ref{cp2}) and $%
\lambda =\frac{15}{l^{2}}$. We obtain 
\begin{equation}
F(r)=\frac{2ml^{2}-r^{6}-3n^{2}r^{4}-3n^{4}r^{2}}{l^{2}(r^{2}+n^{2})^{2}}
\label{712}
\end{equation}%
where the above metric is a solution of the vacuum Einstein field equations
with positive cosmological constant if and only if $2\lambda n^{2}=5$. The
Euclidian section is obtained by analytically continuations $t\rightarrow
i\chi $ and $n\rightarrow in$ (with $l^{2}=6n^{2}$). The NUT solution
corresponds to a one-dimensional fixed-point set of $\partial _{\chi }$
located at $r=n$. The value of the mass parameter is $m_{n}=\frac{-n^{4}}{24}
$ and the periodicity of $\chi $ is $12\pi n$. The bolt solution has a
five-dimensional fixed-point set of $\partial _{\chi }$ located at $r_{b}=%
\frac{kn}{3}$, the value of the mass parameter is given by: 
\begin{equation*}
m_{b}=-\frac{r_{b}^{6}-3n^{3}r_{b}^{4}+3n^{4}r_{b}^{2}}{24n^{2}}
\end{equation*}%
while the periodicity of $\chi $ is $\frac{12\pi n}{k}$. To ensure that $%
r_{b}>n$ we require that $k\geq 4$.

The other possibility is to take $M=M_{1}\times M_{2}$. We could also
consider the possibility of having two different NUT charges $n_{1}$ and $%
n_{2}$ that correspond to the circle fibrations over $M_{1}$ and
respectively $M_{2}$. However, it turns out that in order to have consistent
solutions using the above ansatz we are forced by the field equations to
consider only the cases where $M_{1}=M_{2}$ and moreover (with one exception
- as we shall see below) we must have $n_{1}=n_{2}$. We have then only three
possibilities:

\begin{itemize}
\item $M=S^{2}\times S^{2}$ and the metric is given by: 
\begin{equation}
ds^{2}=-F(r)(dt+A)^{2}+F^{-1}(r)dr^{2}+(r^{2}+n^{2})(d\Omega
_{1}^{2}+d\Omega _{2}^{2})+r^{2}dy^{2}  \label{713}
\end{equation}%
where: 
\begin{eqnarray}
A &=&2n\left( \cos \theta _{1}d\phi _{1}+\cos \theta _{2}d\phi _{2}\right)
\label{714} \\
d\Omega _{i}^{2} &=&d\theta _{i}^{2}+\sin ^{2}\theta _{i}d\phi _{i}^{2}
\label{715}
\end{eqnarray}%
while 
\begin{equation}
F(r)=\frac{4ml^{2}-r^{6}-3n^{2}r^{4}-3n^{4}r^{2}}{l^{2}(r^{2}+n^{2})^{2}}
\label{716}
\end{equation}%
The above metric is a solution of the vacuum Einstein field equations with a
positive cosmological constant $\lambda =\frac{15}{l^{2}}$ if and only if $%
2\lambda n^{2}=5$. The Euclidian section is obtained by the analytic
continuations $t\rightarrow \imath \chi $ and $n\rightarrow in$ (with $%
l^{2}=6n^{2}$). We obtain a solution with negative cosmological constant.
The NUT corresponds to an one-dimensional fixed-point set of $\partial
_{\chi }$ located at $r=n$, the mass parameter is $m_{n}=-\frac{n^{4}}{24}$
and the periodicity of $\chi $ is $12\pi n$. The bolt solution has a
five-dimensional fixed-point set of $\partial _{\chi }$ located at $r_{b}=%
\frac{kn}{3}$, the mass parameter being: 
\begin{equation*}
m_{b}=-\frac{r_{b}^{6}-3n^{2}r_{b}^{4}+3n^{4}r_{b}^{2}}{24n^{2}}
\end{equation*}%
where the periodicity of $\chi $ is $\frac{12\pi n}{k}$. To ensure that $%
r_{b}>n$ we require that $k\geq 4$.

\item $M=T^{2}\times T^{2}$ and the metric is given by: 
\begin{equation}
ds^{2}=-F(r)(dt+A)^{2}+F^{-1}(r)dr^{2}+(r^{2}+n^{2})(d\Omega_{1}^{2}+d%
\Omega_{2}^{2})+r^{2}dy^{2}  \label{717}
\end{equation}%
where: 
\begin{eqnarray}
A &=&2n_{1}\theta _{1}d\phi _{1}+2n_{2}\theta _{2}d\phi _{2}  \label{718} \\
d\Omega_{i}^{2} &=&d\theta _{i}^{2}+d\phi _{i}^{2}  \label{719}
\end{eqnarray}%
while 
\begin{equation}
F(r)=\frac{4m}{(r^{2}+n_{1}^{2})(r^{2}+n_{2}^{2})}  \label{720}
\end{equation}%
The above metric is a solution of the vacuum Einstein field equations for
any values of the parameters $n_{1}$ and $n_{2}$. We obtain the Euclidian
section by analytic continuation $t\rightarrow i\chi $ and $n_{j}\rightarrow
in_{j}$, $j=0,1$. Notice that $F_{E}(r)$ does not vanish for any value of $r$
and that there is a curvature singularity located at $r=n$, where $n$ is the
maximum value of $n_{1}$ and $n_{2}$.

\item $M=H^{2}\times H^{2}$ and the metric is given by: 
\begin{equation}
ds^{2}=-F(r)(dt+A)^{2}+F^{-1}(r)dr^{2}+(r^{2}+n^{2})(d\Sigma
_{1}^{2}+d\Sigma _{2}^{2})+r^{2}dy^{2}  \label{721}
\end{equation}%
where: 
\begin{eqnarray}
A &=&2n\cosh \theta _{1}d\phi _{1}+2n\cosh \theta _{2}d\phi _{2}  \label{722}
\\
d\Sigma _{i}^{2} &=&d\theta _{i}^{2}+\sinh ^{2}\theta _{i}d\phi _{i}^{2}
\label{723}
\end{eqnarray}%
while 
\begin{equation}
F(r)=\frac{4ml^{2}+r^{6}+3n^{2}r^{4}+3n^{4}r^{2}}{l^{2}(r^{2}+n^{2})^{2}}
\label{724}
\end{equation}%
The above metric will be a solution of the vacuum Einstein field equations
with a negative cosmological constant $\lambda =-\frac{15}{l^{2}}$ if and
only if $2\lambda n^{2}=-5$. The Euclidian section is obtained by analytic
continuation $t\rightarrow i\chi $ and $n\rightarrow in$, with $l^{2}=6n^{2}$%
. The NUT solution corresponds to an one-dimensional fixed-point set of $%
\partial _{\chi }$ located at $r=n$, with the mass parameter $m_{n}=\frac{%
n^{4}}{24}$ and the periodicity of $\chi $ being $12\pi n$. The bolt
solution has a five-dimensional fixed-point set of $\partial _{\chi }$
located at $r_{b}=\frac{kn}{3}$, the mass parameter: 
\begin{equation*}
m_{b}=\frac{r_{b}^{6}-3n^{2}r_{b}^{4}+3n^{4}r_{b}^{2}}{24n^{2}}
\end{equation*}%
and the periodicity of $\chi $ is $\frac{12\pi n}{k}$, $k$ being an integer.
To ensure that $r_{b}>n$ we require that $k\geq 4$.
\end{itemize}

\section{Conclusions}

We have considered here higher dimensional solutions of the vacuum Einstein
field equations with (or without) cosmological constant that are locally
asymptotically de Sitter, anti-de Sitter or flat. These solutions are
constructed as circle fibrations over even dimensional spaces that can be in
general products of Einstein-K$\ddot{a}$hler spaces. The novelty of our
solutions is that, by associating a NUT charge $n$ with every such circle
fibration over a factor of the base space we have obtained higher
dimensional generalizations of the Taub-NUT spaces that can have quite
generally multiple NUT charges. We have also generalized the Taub-NUT ansatz
to the case of odd-dimensional spacetimes and we have explicitly constructed
the form of the Taub-NUT solutions in five and seven dimensions.

We found that depending on the specific form of the base factors on which we
construct the circle fibration we can have cases in which the cosmological
constant can have either sign. Another interesting characteristic of our
solutions is the unexpected relation between the NUT charges and the
cosmological constant. We found that quite generally, in dimensions higher
than four, it is impossible to make either all the NUT charges or the
cosmological constant vanish independently. In other words our solutions do
not have non-trivial asymptotically flat NUT-charged limits that are
obtained by simply taking the limit $\lambda\rightarrow 0$.

Even if, for space reasons, we have given the solutions up to seven
dimensions, our solutions can be generalized in an obvious way for
dimensions higher than seven. For instance, in eight dimensions, the full
base space is six-dimensional and in general we can factorize it in product
manifolds constructed out of combinations of two-dimensional spheres, tori
or hyperboloids. For each such two dimensional factor one can associate a
NUT charge in the final circle fibration. We can also consider the base
factor as a product of a four-dimensional manifold (like $CP^2$ or more
generally an Einstein manifold of constant curvature) with a two dimensional
one (which could be a sphere, a torus or a hyperboloid). For every factor
that is an Einstein-K$\ddot{a}$hler manifold we can associate a NUT charge
parameter.

In our work we have given the Lorentzian form of the solutions however, in
order to understand the singularity structure of these spaces we have
concentrated mainly on the their Euclidian sections. In most of the cases
the Euclidean section is simply obtained using the analytic continuations $%
t\rightarrow it$ and $n_{j}\rightarrow in_{j}$. However, given the special
relationship between the NUT charges and the cosmological constant, in some
cases we also have to analytically continue $l\rightarrow il$ in order to
obtain a consistent solution with Euclidian signature. When continuing back
the solutions to Lorentzian signature the roots of the function $F(r)$ will
give the location of the chronology horizons since across these horizons $%
F(r)$ will change the sign and the coordinate $r$ changes from spacelike to
timelike and vice-versa.

Leaving the study of the thermodynamic properties of these solutions for
future work, it is worth mentioning that our solutions can be used as
test-grounds for the AdS/CFT correspondence. For our solutions the boundary
is generically a circle fibration over base spaces that can have exotic
topologies and this could shed some light on the study of the conformal
field theories ($CFT$) on backgrounds with such exotic topologies. In
particular one could be able to understand the thermodynamic phase structure
of such conformal field theories by working out the corresponding phase
structure for the our supergravity solutions in the bulk.

\bigskip

This work was supported by the Natural Sciences and Engineering Council of
Canada.

\end{document}